\documentclass{article}

\RequirePackage{graphicx}
\usepackage{graphicx} 
\usepackage[utf8]{inputenc}
\usepackage{physics}
\usepackage{graphicx} 
\usepackage{epsfig}

\usepackage[english]{babel}

\usepackage[letterpaper,top=2cm,bottom=2cm,left=3cm,right=3cm,marginparwidth=1.75cm]{geometry}

\usepackage{amsmath}
\usepackage{url}

\usepackage[colorlinks=true, allcolors=blue]{hyperref}

\title{\bf{Women in STEM: Interview with Iris Abt}}
\author{Elisabetta Gallo$^{1,2}$, Henriette Ullmann$^{2}$ \\
$^1$Deutsches Elektronen-Synchroton DESY, 
Notkestr. 85, 22607 Hamburg, Germany \\
$^2$Universit{\"a}t Hamburg, Germany}

\usepackage[style=numeric,sorting=none]{biblatex}
\usepackage{comment}
\begin{document}

\maketitle
\tableofcontents

\begin{abstract}
  We interviewed Iris Abt, who studied in Germany, being the only woman finishing studies in her course during that year. She then started her career in neutrino physics, moved to SLAC at the times of SLC, came back to Europe shaping part of the HERA program and made studies on germanium detectors.
\end{abstract}

\section{Introduction}

When we asked Iris Abt whether she wanted to make an interview with us, she answered that there was no point, that she was always in the wrong place at the wrong time and that she did not accomplish anything. 
She also said that her experience in HEP had not been entirely
happy.
Our answer was that we wanted to hear from her anyway. 
Iris Abt studied in Hamburg with a PhD in the CHARM (neutrino) group and moved to SLAC for her postdoc. She came back to Europe becoming a member of the senior (permanent) staff at the the 
Max-Planck-Institute of Physics (MPI) in Munich from where she  retired in 2024. At the MPI, she became a project manager for the HERA program.
She also joined the quest to observe neutrinoless double-beta decay,
for which she also started an R\&D program on germanium detectors. We cover this long journey including also some anecdotes.

\section{Studying in Hamburg}

Iris Abt studied at the University of Hamburg in Germany. At the time, studying in Hamburg was very flexible. There was a lot of freedom in organizing the own schedule and in choosing the sequence and selection of lectures to obtain a Diploma. The only strict requirement was to take the lectures Physik~I to Physik~IV and Quantum Mechanics~I, Electrodynamics~I, Thermodynamics~I. All this freedom allowed her to study
in parallel mathematics and to get a Diploma in mathematics, learning  numerics very early. Her specialties were unfolding methods and integral equations, a very useful portfolio for her later research in physics. She particularly admired Prof.\,Lothar Collatz, the co-supervisor of her Diploma thesis, for his way to lecture and talk to students. Interestingly, in mathematics there were about 60\% women, while in physics there were 350 students and only three women starting in October 1976. 
Iris was the only one of the three women who graduated as a physicist.
The other two were discouraged by remarks and circumstances.

Iris wanted to pursue her second Diploma thesis in theory with a subject on baryogenesis. But the professor she asked, learning that she already had a Diploma in mathematics, sent her to the dean, Prof.\,Peter St\"ahelin, to permit the direct quest for a PhD in physics. So the dean interviewed Iris, talked shortly with her supervisor in mathematics, the {\emph{Diploma father}}, and told her that they decided that she was {\emph{too intelligent to become a theoretician}}. On the spot, he offered her a position in the neutrino group at Hamburg. At that time, PhD students in theory had to finance themselves, while in experimental physics half of a position was offered. Iris answered that she needed some time to think about it until after her vacation. At first, Prof.\,St\"ahelin agreed but
then he tracked Iris down in France where she was camping with a friend as he needed a decision. Iris was so impressed by the dean tracking her to a remote campground that she accepted the position. She then started her research in neutrino-nucleon scattering with the CHARM experiment at CERN in October 1981.

Iris was the first PhD female graduate student in that group and she made it clear on the second day that she was not there to cook coffee.
It took her 4 years, 7 months and 25 days to graduate, spending half of her PhD time at CERN. Her thesis was about the inelasticity distributions of neutrino and antineutrino nucleon interactions.
The data was also used to determine the ratio of neutral current (NC) to charged current (CC) cross section, yielding a very precise determination of the Weinberg angle~\cite{CHARM:1987pwr}.
Her time as a graduate student was not a smooth path but she is proud to have persevered and have graduated.

She is proud of her first publication describing the calibration of the CERN narrow-band neutrino beam~\cite{Abt:1984fh}.
The flux of neutrinos was determined by measuring the muons produced together with the neutrinos.
For each pulse of the SPS about $10^{10}$ muons were produced and their flux was monitored in the so-called muon shield, a 400\,m steel wall with slots for detectors placed at specific places in this shield. At that time, solid state silicon detectors were used for the measurement and their signals needed to be calibrated. Nuclear emulsions were placed on covering 10 of the silicon detectors and the SPS delivered one pulse in this configuration. Afterwards, the nuclear emulsions were taken out and Iris drove them to Amsterdam, where they were scanned to identify about 1000 tracks per emulsion. From the number of muon tracks per unit area, the neutrino flux was calculated. 

Before the new calibration, there was a 25\% difference between the total neutrino-nucleon cross-section as measured at FNAL and at CERN. The new calibration moved the CERN result halfway towards the FNAL result. 
While the CHARM Collaboration did not publish this calibration, Iris published together with the head of the Amsterdam scanning team, Bob Jongejans.  Later, CERN found out that the beam-current transformers had not been properly grounded and FNAL found out they had used the wrong target mass. The two results became compatible and this story proves once more how
important independent measurements are.

It was not easy to be a woman in HEP at the time. The person responsible for the solid state detectors was Eric Heine and
as Iris was to be responsible for the nuclear emulsions and the analysis, one day Eric wanted Iris to go down with him into the shield and help mounting the emulsions. 
However the CERN {\emph{pompiers}} denied her access to the shield. As a woman she would probably faint climbing down the ladder and they would have to carry her up. Iris was ready to give up, but Eric insisted that she would go down with him. It took first the division head and then even the director general himself  to ask the head of the fire brigade to order his men to let Iris down to the neutrino shield, the whole took three hours.
 Iris is still grateful to Eric who was very supportive during the whole procedure. She wanted him to be an author of the paper but he declined of account he didn't do enough. 

Iris spent time at CERN from 1982 to 1984.
These were very exciting times at CERN. There was always a rumor about something new in the cafeteria. 
{\emph{We were always expecting that something would be discovered}}, says Iris. 
She was there when Carlo Rubbia showed the first $W$
and $Z$ events. 
She was also there when the Mont Blanc experiment presented three events that hinted at proton decay. Iris was a bit late and sat on the balcony. Also sitting there was a theorist, Alvaro de Rujula, 
who seemed unusually happy. After the three events looking like proton decay had been presented in detail, he raised his arm and asked whether {\emph{neutrino background had been considered?}}. The speaker as well as the audience froze. You could have heard the proverbial pin drop. It was the only time in her life that Iris had a telepathic experience. Everybody in the packed auditorium thought {\emph{{Oh no, we forgot about that.}} That was the start of neutrino astroparticle physics.

\section{The SLAC period and who is that guy asking so many questions?}

After the PhD, Iris went to SLAC where the linear $e^+ e^-$ collider SLC was being built. She really wanted to work on hardware and especially on the SLD experiment as it was still in an early construction phase, offering a great opportunity.
She had job interviews at Caltech, SLAC and Santa Cruz. 

Her first stop at Caltech provided another anecdote.  She gave an overview talk on neutrino physics, 
at the time CHARM had a result indicating the violation of lepton universality (LFU) through the production of the number of electron versus muon neutrinos in a beam dump. During the talk there was this old guy in the audience constantly asking questions. Iris was a bit annoyed as she had to permanently adjust her timing. She thought that every faculty had a strange guy attending seminars. He was especially interested in the LFU result and after the end of the talk, he asked again whether Iris had an explanation.
She said that she was convinced it was an experimental problem but
he declared it was interesting and he would think about it.
Iris had enough of it and answered {\emph{If you haven't anything better to do.}} After the question time, he joined Iris when she tried to
sort her transparencies. She wondered why the organizers didn't rescue
her but the discussion became more and more interesting and focused on
vacuum polarization. Some student came up and remarked that it was all explained by Feynman diagrams. Iris thought "what nonsense" and said
that {\emph{Feynman diagrams are a way to calculate path integrals but explain nothing}}. The old guy agreed and they continued their discussion.
The organizer then interrupted their conversation after an hour as it was time for dinner. The old guy apologized and said he could not join. At the dinner, the organizer remarked {\emph{It is always nice when Feynman comes to the seminars and asks questions}}. Iris wondered who besides the old guy asked questions and who was Feynman. Then the organizer said that {\emph{Feynman really seemed to like you. I have never seen him talk that long to a speaker}}. Iris thought {\emph{Oh my lord, I should run away}}. She was very surprised when she learned
that Caltech wanted to hire her after all.

In the end, she preferred to join the SLD collaboration based at SLAC because she wanted to be where the detector was built. She got hired, even though she had basically no hardware experience. The speaker
of SLD, Marty Breidenbach, asked her why he should hire someone without
experience. She answered \emph{Because I learn faster than anybody else.} Later they told Iris that this got her the job. 

Iris really had a wonderful time in California, and the SLAC attitude towards women was much more open. Iris learned that {\emph{It were women who drove the wagons across the plains and so men in the west have always appreciated strong women}}. 
On her second or third day, she and two colleagues transported 
some equipment to a test site. After the first load she was asked to watch the equipment to prevent it being stolen while the men would get the rest. And while she was waiting, she could install two NIM crates in the rack. This was a joke! No human being can install a NIM crate
with power supply by him/herself. Iris was surprised but pleased by the joke. 
She found two cryotechs who willingly helped her
and hid before the colleagues came back. They were stunned by the
site of Iris leaning against the rack, smiling and stating it had been easy.
It ended in a lot of laughter and Iris knew she would be comfortable and happy at SLAC.

In just two years, she developed from a simple postdoc performing tests of the electronics of the Liquid Argon calorimeter to a person
entrusted with leadership during its final construction phase. She became a {\emph{scheduling scientist}}, responsible for everything that was happening in the hall, including safety. 
Iris was very lucky, as SLD had Marty Breidenbach as a leader she
could truly respect. He would listen to the people actually doing the work independent of their status. He was also able to take decisions.
Iris says \emph{Many so called leaders in physics lack this ability}. He was a role model for her as he was combining leadership with technical expertise.  Two other noteworthy people who were part of the calorimeter construction and testing team, instrumental for SLD, were Eric Vella and Alan Honma. The latter became very important for the CMS vertex detector. These two men and Iris became friends for life. 
Iris says {\emph{If you work together on a detector for days and nights and really have to trust each other  you develop a strong bond for life}}.

SLD was a modern detector with a CCD-based vertex detector, drift chambers for tracking and Cherenkov Ring Imaging detectors for particle identification. The LAr calorimeter~\cite{Axen:1990dz} was placed inside the magnetic coil and augmented by a warm iron calorimeter and muon detector outside. 
{\emph{At the time I joined SLD, the physics of the $Z^0$ was expected to be exciting. Everybody was thinking there will be at least one or two discovery, like toponium, technicolour, Supersymmetry...}} says Iris. However, 
the SLC was late and never reached the promised luminosity.
Before SLD was finished, the MARK II detector took the first data at the SLC. MARK II had to take 80 hours of shift per $Z^0$.
The collaboration managed to collect
about a 1000 $Z^0$ events and publish the measurement of the number of
light neutrinos as $2.8\pm0.6$~\cite{Abrams:1989yk} in November 1989.

When SLD moved into the SLC interaction region, LEP was running.
With SLC providing less luminosity than CERN, SLD could only compete with the four LEP detectors because it was an excellent detector and because SLC had a $70-80\%$ polarized electron beam. 
Thus it could precisely measure the so-called 
left-right asymmetry, one special method to determine the Weinberg angle~\cite{SLD:2000leq},
\begin{equation}
    A^0_{LR} = 0.15056 \pm 0.00239.
\end{equation}
This value is still the most precise value available on $A_{LR}$.
SLD was a wonderful detector (Fig.~\ref{fig:LAC}).
It operated from 1991 to 1998 and collected about 500\,k $Z^0$ events and
the SLD collaboration was able to publish competitive papers in several fields.

\begin{figure}
\begin{center}
\includegraphics[angle=-90,width=1.0\textwidth]{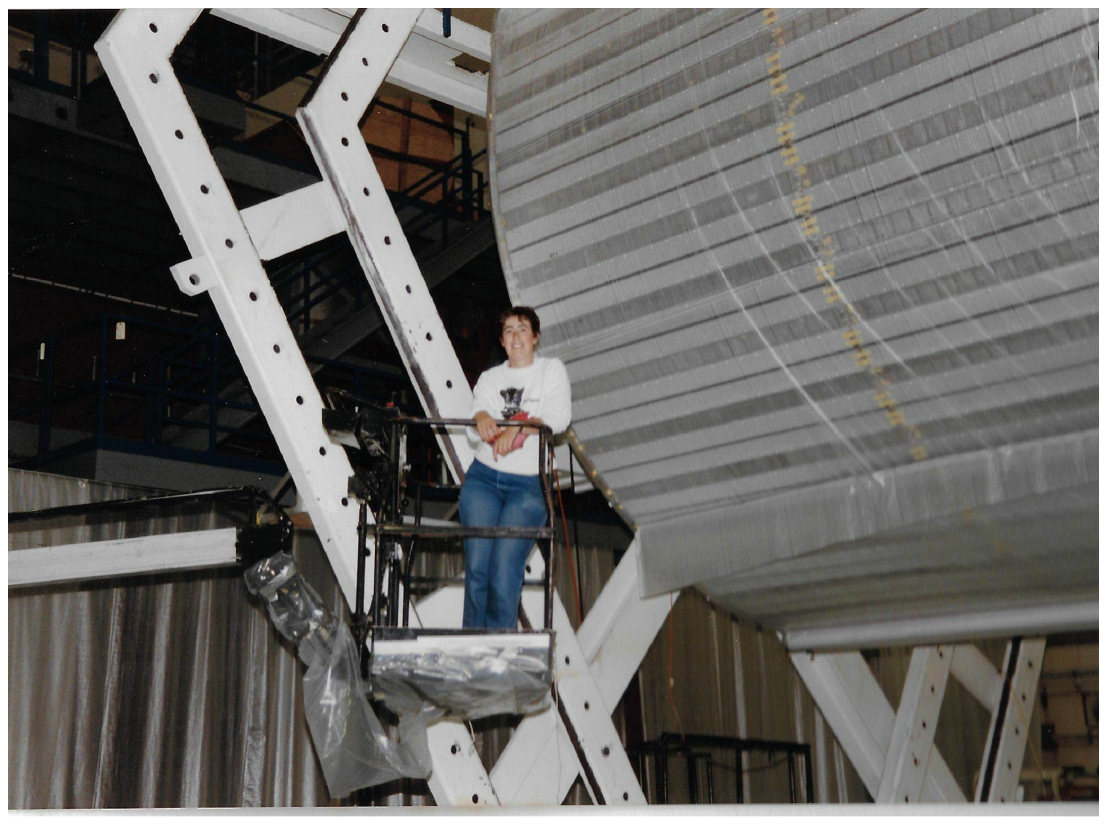}
\end{center}
\vspace{-2cm}
        \caption{Iris Abt next to the SLD liquid Argon Calorimeter
        during the last inspection before it was mounted into its cryostat [Credit picture: Iris Abt, June 1988 ] } 
      \label{fig:LAC}
    \end{figure}

However, the future of high energy physics at SLAC was uncertain and
after six years, Iris left and worked for the University of California
Davis for a year on the H1 experiment at DESY in Hamburg. But then
an opportunity to go back to Germany came up.
She became part of the senior (permanent) staff at the Max-Planck-Institute for physics  (MPI) in Munich. She was hired to lead the ALEPH group, 
responsible, among other things, for the repair of the modules of the silicon vertex detector. After a year, however, she was asked
to start and lead a group to build the modules of the
silicon vertex detector of the HERA-B detector at DESY in Hamburg. 

\section{The HERA-B period} 

 The HERA-B experiment was a fixed target experiment to study heavy flavor physics, exploiting the 920 GeV proton beam of HERA colliding with a wire target with three different nuclei (C, W, Ti). It was a large-aperture spectrometer that had to cope with high rates and extremely high radiation levels for the detectors.
 The vertex detector was at the forefront of technology and
 the solutions for HERA-B later helped to build detectors to be operated in similar radiation loads at the LHC.
 
The MPI group joined in 1994 when  HERA-B was still in the design
phase. Iris was very anxious about the aggressive schedule advertised by the HERA-B management and DESY as well as the general budget.
The HERA-B vertex detector was the first silicon vertex detector designed for a high radiation environment and with individual detectors
being exposed to radiation inhomogenously.
It was based on eight layers, each with four pairs of double-sided detectors and had 147456 analogue channels operated in vacuum. Iris' task in HERA-B was to design, produce and install the five kinds of silicon modules in the roman pots on the HERA proton beamline. The infrastructure for the modules was built by the MPIK Heidelberg group headed by Karl Tasso Kn\"opfle.  Some pictures from the assembly of the vertex detector are shown in Fig.\,\ref{fig:HERA-B}
\vspace{-0.5cm}
\begin{figure} [h]
\includegraphics[angle=0,width=0.5\textwidth]{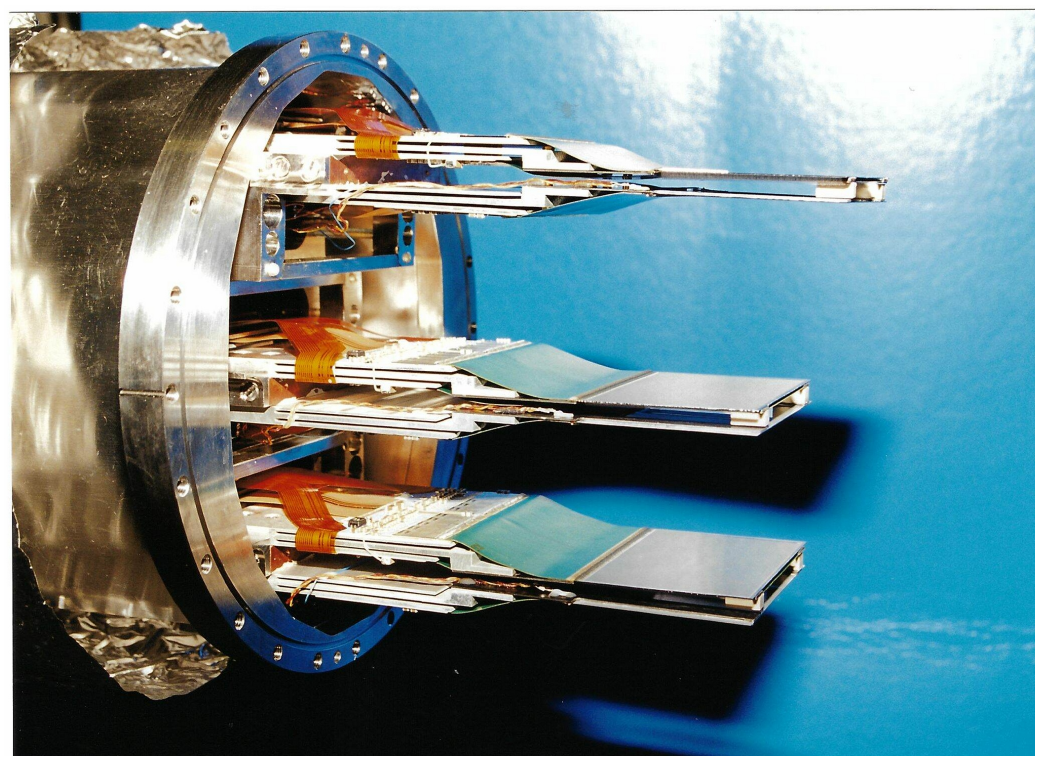}
\hspace{-1cm}
\includegraphics[angle=0,width=0.5\textwidth]{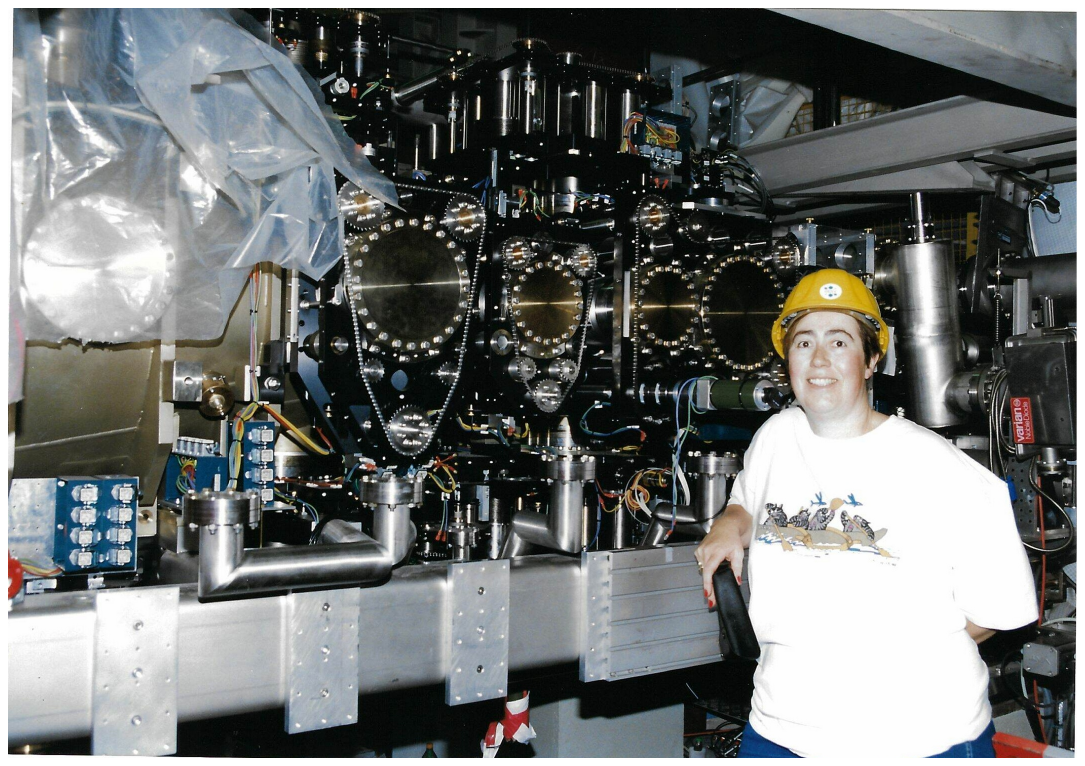}
\end{figure}
\vspace{-6cm}
\begin{figure} [h]
\includegraphics[angle=0,width=0.5\textwidth]{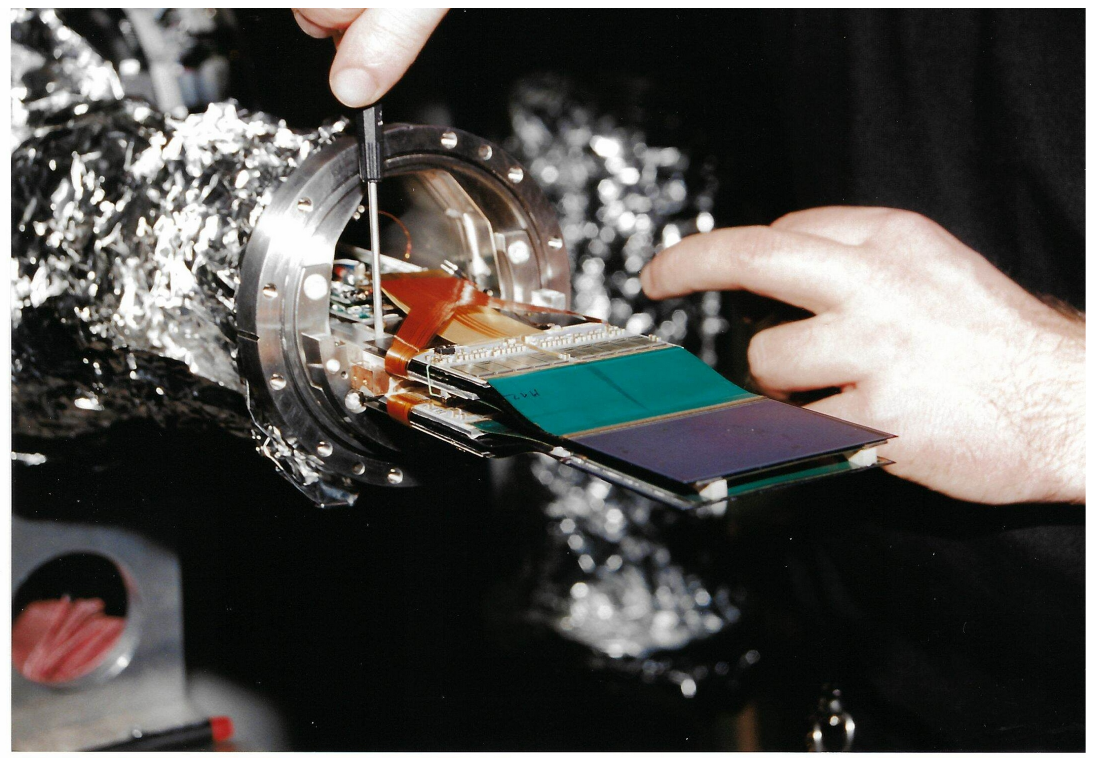}
\vspace{-6cm}
        \caption{Six detector modules forming one quadrant of the first three layers of the vertex detector; Iris Abt standing in fron of the HERA-B vertex detector vacuum vessel;
        One detector module of layer 5 being mounted 
        [Credit pictures: Iris Abt 1988/89] } 
      \label{fig:HERA-B}
\end{figure}

The Max-Planck-Society provided enough budget and people for the project to enable the two MPIs to complete the project according 
to the ambitious schedule.
However, HERA-B as a whole experienced many unexpected difficulties and delays.
It was originally planned to start ahead of the $B$-factories in Japan and the USA, but
 it was never really completed before the long HERA shutdown
to upgrade the $ep$ collider (2000--2003).

In spite of this, the vertex detector worked extremely well and withstood the harsh
environment. Many of the technical solutions were taken over by later experiments, like the flexible glue used to fix the silicon detectors
to the carbon support structures.
The vertex detector had enough redundancy to reconstruct standalone tracks. This was done by employing a cellular automaton~\cite{Abt:2002he}.   
As many of the components, especially the trackers, had problems and
time was running out, Iris insisted on data to be taken without
these components and without trigger. The recorded 
minimum bias events were basis for some basic analysis
to allow all of her students to graduate on simple subjects like event multiplicities or vertex distributions and resolution.

In general, it was very difficult for students in HERA-B to graduate.
The trigger for B physics did not work and the detector as a whole
did not reach the expected performance. 
At the end of the HERA I period in the year 2000, there was some interest of the cosmic ray community to get cross section values from the proton collisions with the different target materials. HERA-B would have liked to run with only the proton beam for one month at the end of HERA I to provide these measurements, but this was not supported
and at that point, the HERA-B collaboration basically started to
dissolve.
The Extended Scientific Council of the DESY laboratory, seeing the poor performance compared to the $B$-factories and the lack of manpower, recommended {\emph{an orderly termination in the near future}}~\cite{LohrmannSoeding} and to terminate data taking for HERA-B  around end of 2002. In 2002, the remnant groups of the collaboration  presented a revised physics plan and took data for four months in 2002-2003, collecting 300k $J/\Psi$ events and 210M minimum bias events. Soon after, the collaboration started dissolving completely. However, scientists gained a lot of experience and it is fair to say that all the technical expertise accumulated for HERA-B was crucial for the success of LHCb. 

\section{The ZEUS and H1 combined data}

In 2002, Iris was transferred into the MPI department headed by 
a new director, Allen Caldwell. Her first task was to design a detector for forward physics for his HERA~3 project. While HERA~3 was not approved at DESY, Iris presented the detector design at a conference in New York, where one of the subjects was a possible electron ion collider (EIC) in the US. Some of her concepts were later incorporated into a detector design for the EIC, for instance the geometry of the vertex detectors with a short barrel and a large number of disks in the forward region. She also discussed possibilities to build a HERA~3-like detector at Brookhaven with their machine group. Unfortunately, the machine lattice could not be modified to allow for a large enough magnet free interaction region.
As a member of Allen Caldwell's department, she then joined the ZEUS experiment at HERA, one of the two general purpose detectors measuring deep inelastic scattering.

Her first contribution was to point out in ZEUS physics meetings that  data and Monte Carlo control distributions should be looked at not only in logarithmic but also in linear scale. That was enough for the ZEUS management to make her a coordinator of the high $Q^2$ group.
In 2008, the spokesperson at the time, Tobias Haas, asked her to become the physics coordinator (ZEUS physics chair). 
She guided the publication of 18~papers that year, a record number for ZEUS. As a result she was asked to stay and she has been covering that role since then. After the end of HERA data-taking in 2007, there were a lot of analyses ready to be published, it just needed a better organization of the paper publishing process in ZEUS, where Iris turned out to excel.

\begin{figure}
\vspace{-5cm}
\begin{center}
\includegraphics[width=0.9\textwidth]{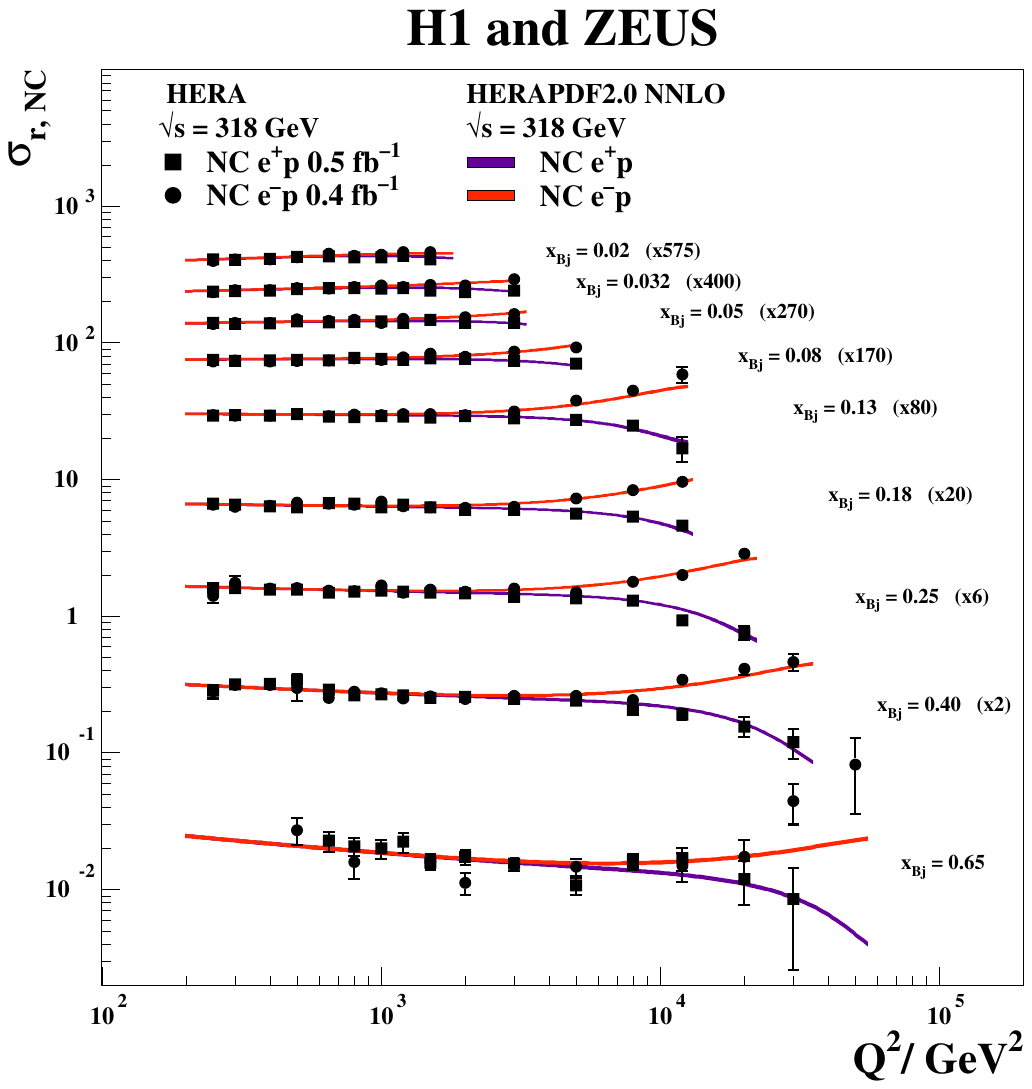}
\end{center}
        \caption{Representative points in the combined HERA data for the inclusive NC  and the predictions of the QCD fit HERAPDF2.0 NNLO. The bands represent the total uncertainties on the fits.  (from~\cite{H1:2015ubc}). } 
      \label{fig:Combined_H1ZEUS}
    \end{figure}

Her biggest achievement was certainly the writing of the combined legacy H1+ZEUS paper on inclusive cross sections and combined QCD fits. It was published in 2015 and presented a combination of neutral current (NC) and charged current (CC) cross sections, using all  $e^\pm p$ data collected in HERA I and HERA II, corresponding to about 1 fb$^{-1}$ of integrated luminosity~\cite{H1:2015ubc}. The data cover  wide ranges in $Q^2$  and Bjorken $x$, $0.045 <Q^2< 50,000$ GeV$^2$ and $6 \times 10^{-6} < x < 0.65$. The paper also presented a perturbative QCD analysis of the data which was performed to extract proton parton densities (PDFs) in the HERA parameterization. Figure~\ref{fig:Combined_H1ZEUS} reports the NC reduced cross sections as a function of $Q^2$ at fixed $x$, showing the very large range covered. The HERAPDF2.0 fits at NNLO provide a very good description over the whole range and parity violation at high $Q^2$ is evident from the difference between the $e^+p$ and $e^-p$ cross sections.  The importance of these data 
cannot be stressed enough  as they are the only set of consistent $ep$ collider data and are used to extract PDFs by the various fitting groups. They have been crucial to predict precise cross sections at the LHC. Everybody admitted that the paper would never have been published without Iris. Iris mentions Mandy Cooper-Sarkar, Katarzyna Wichmann and Matthew Wing as fundamental people working on those QCD fits during those years, including a seminal paper on a combined PDF fit together with a limit on the radius of the quark ~\cite{ZEUS:2016ygq}. A combined fit is the more appropriate way as, if BSM physics effects existed in the HERA data the "normal" PDF sets are biased. The resulting 95\% confidence upper limit on the effective quark radius was $0.43 \times 10^{-16}$ cm and Iris won a bet of twelve bottles of champagne from Thomas Hebbeker on the fact that substructure of quarks had not been found 20 years after their time together as graduate students in Hamburg.

ZEUS had stopped data-taking in 2007, but it had a very reasonable way to do data preservation, based on ntuples~\cite{Antcheva:2009zz}, thanks to the visionary work of Achim Geiser. This allowed to react to new developments that came later and the data attracted several groups, that joined ZEUS to produce a couple of very interesting papers. The first paper was triggered by a group from the heavy-ion community at ALICE, who had an interest to study two-particle correlations for high track multiplicities in $ep$ collisions~\cite{ZEUS:2019jya} after a ridge structure was observed in $pA$ and $pp$ collisions at RHIC and LHC. The correlations in $ep$ did not indicate any kind of collective behavior and could be described by perturbative QCD Monte Carlo simulations.  Another paper was a search for the violation of isotropy in our universe~\cite{ZEUS:2022msi}.  As rotations and boosts do not commute, the violation of rotation invariance implies the violation of boost invariance and vice versa. A vector field, instead of the scalar Higgs field, could create such anisotropy and cause a dependence of the $ep$ cross section on time. No time dependence was observed and interesting limits were set on coefficients in an Effective Field Theory with CPT violation.
Even though Iris was not very interested in the beginning, she became quite enthusiastic about this result.

\section{$0\nu\beta\beta$ and germanium}

Soon after HERA-3 was not approved, Iris got a call from Karl-Tasso-Kn\"opfle whether she would not like to join a project later
called GERmanium Detector Array (GERDA)~\cite{Abt:2007ba} to search for neutrinoless double-beta decay ($0\nu\beta\beta$) in $^{76}$germanium. 
The name GERDA was chosen in a meeting where all suggestions were voted on. Everybody voted for their own suggestions but two people, one of them Iris, had a mother called Gerda. That made the difference.

Iris wanted a small project for herself and
her postdoc Xiang Liu who had helped with HERA~3. 
So she first agreed to do some mechanical engineering for the GERDA detector array. 
However, it became a very large project, including the fundamental research on germanium detectors done by the newly created GeDet group. 

This fundamental research was interesting in so far as the knowledge about germanium itself and germanium detectors was much more limited than Iris had expected. While silicon properties are well studied, even the mobilities in germanium are not well known.
The group developed special test detectors together with the Lingolsheim detector manufacturing site of Mirion (formerly Canberra)
technologies and special test stands to extract some of these fundamental properties.
 After the GERDA collaboration finished its program without finding
$0\nu\beta\beta$, 
it formed the LEGEND collaboration together with its US counterpart, the MAJORANA collaboration, and some other groups in 2016.
It was very clear from the beginning that pushing the search for
$0\nu\beta\beta$ further, a better understanding of germanium was elementary.
It was natural for the GeDet group to became part of this effort.
Iris was also involved in developing the rules governing the collaboration and 
she was elected to lead the LEGEND collaboration board for a year in 2021.

\begin{figure}
\begin{center}
\includegraphics[width=0.45\textwidth]{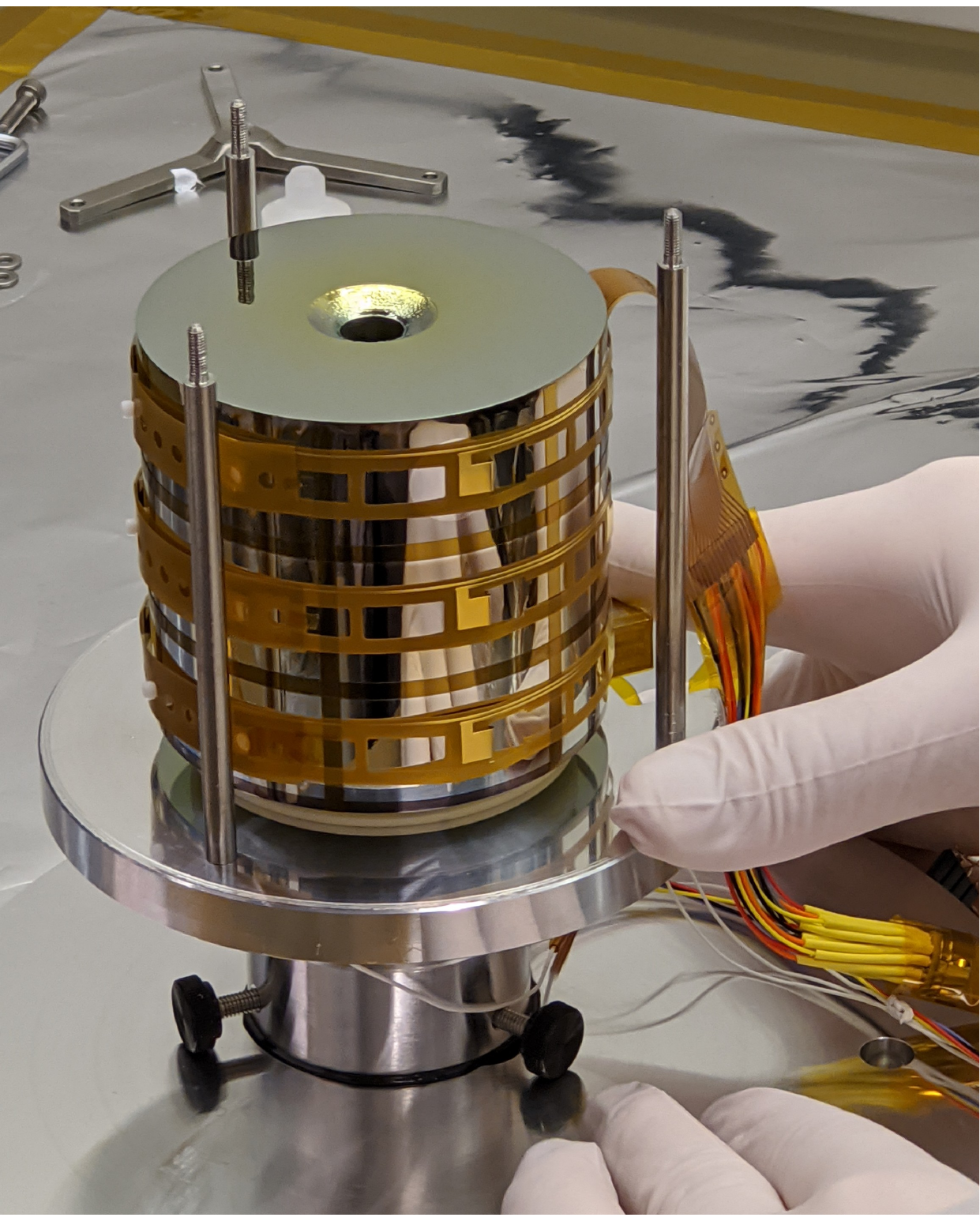}
\end{center}
\vspace{-0.5cm}
        \caption{An 18-fold segmented Germanium detector taken outside its cryostat for a modification of its cabling.  [Credit: Iris Abt, 2019] } 
      \label{fig:germanium}
    \end{figure}

The work on germanium detectors had started in 2004 and had lead to interesting results about the physics inside these detectors. 
The last 6 years before Iris' retirement in 2024, the GeDet group worked
according to a 5--year plan to measure fundamental parameters of germanium in large scale ($> 100$\,cm$^3$) detectors (Fig.~\ref{fig:germanium}).
This focused on surface effects, pulse shape formation, depletion, impurities and charge carrier mobility~\cite{Abt:2022odr, Abt:2022vwo}.
She had a final group of very good graduate students, Martin Schuster, Lukas Hauertmann and Felix Hagemann,
an excellent IT expert (Oliver Schulz) and her previous postdoc Xiang Liu, who had returned after some years as an assistant professor in Shanghai.
The students and Oliver Schulz 
proposed to write a new simulation package for germanium detectors using modern computing techniques and the acquired knowledge about
germanium.
The resulting open source package $SolidStateDetectors.jl$~\cite{Abt:2021mzq} (SSD) calculates the electric field, simulates charge carrier drift and produces pulse shapes in germanium detectors. It is now widely used in the field. 
SSD is an essential tool to better understand germanium detectors,
especially the interplay of 
impurities and charge carrier mobilities in forming the pulses obtained from
germanium detectors, see Fig.\,\ref{fig:epot-components}.

\begin{figure}[!tb]
    \centering
    \includegraphics[width=1.0\textwidth]{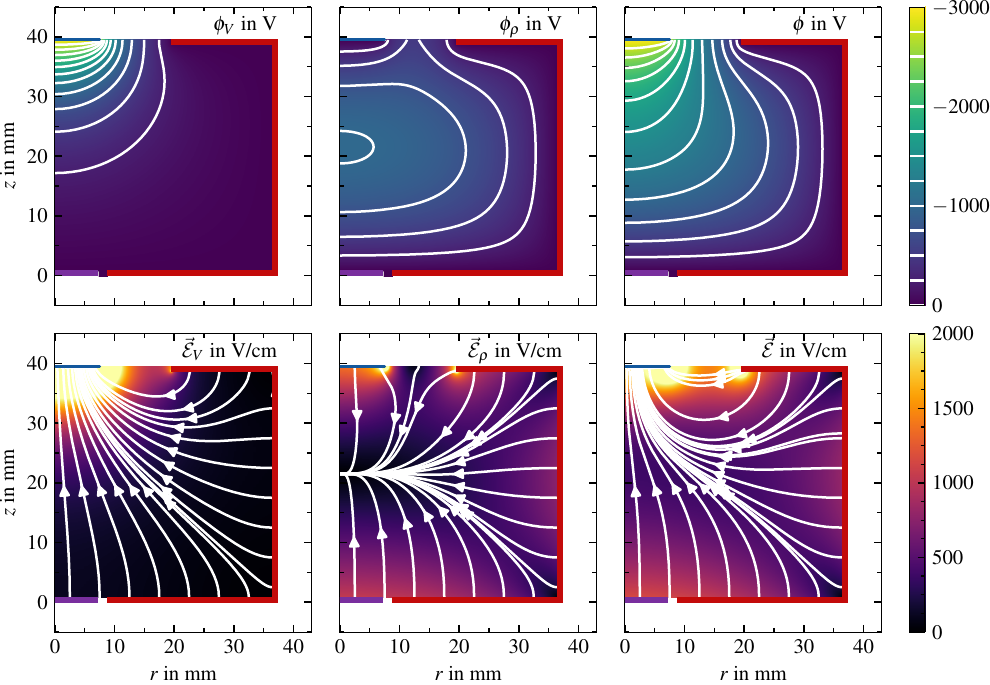} 
    \caption{A cut of half the detector is shown. Top row: electric potential in the $r$-$z$-plane of a point-contact p-type germanium detector resulting from (left) applying just the potentials to the contacts, $\phi_V$, (center) just the charge density from ionized impurities, $\phi_\rho$ and (right) total electric potential, $\phi$. The white lines depict equipotential lines at integer multiples of $-250$\,V.
    Bottom row: corresponding electrical fields. The white lines represent field lines  guiding the trajectories of charge carriers in the total field~\cite{abt:bubbles}}.
    \label{fig:epot-components}
\end{figure}

The group managed for the first time to image the undepleted volumes of a germanium detector and to deduce the radial dependence of
detector impurities~\cite{abt:bubbles}. A publication presenting the first measurement of charge carrier mobilities in a large ( $> 100$\, cm$^3$) cylindrical detector is under preparation~\cite{Hagemann2024} .

\section{Iris and China}
Even as a teenager, Iris was fascinated by China. It was a closed country in the 1960s and her interest was sparked by movies like
"The Sand Pebbles, 1966" (Kanonenboot am Yangtse-Kiang).
So she knew nothing, learned nothing in school and never even dreamed, she would get there.

In 1995, there was a Lepton Photon conference in Beijing and this
was her chance to go there. She explored Beijing and took part in a
pre-conference tour to Tibet and a post-conference river cruise down the not yet dammed Yangtse river. It was a very interesting trip.
Travel to not easily accessible places is one of the benefits of working in international science.
Iris did not expect to travel to China again.

However, in 2019, her colleague and friend 
Xiang Liu became an assistant professor at Shanghai Jiao Tong 
and he discovered the Sino-German-Center for sciences.
They support cooperation between Chinese and German groups.
It required quite some paperwork, but the Chinese elite universities Tsinghua in Beijing and Shanghai Jiao Tong, together with the MPI Munich  and T\"ubingen university formed a cooperation. The first meeting happened in March 2011 in Beijing. The subject
of the collaboration was germanium detectors and large scale germanium based experiments to search for $0\nu\beta\beta$ or dark matter.
A visit to the new first phase of the Chinese JinPing underground Laboratory (CJPL) close to Xichang in Sichuan was also included.

It was a very informative and pleasant experience and after some more paperwork, three very good symposia about germanium detectors and applications in fundamental science followed. They were organized by T\"ubingen (2013), Tsinghua (2014) and the MPI Munich (2015). The cooperation with China also led to Tsinghua university and some other Chinese groups joining the LEGEND collaboration. 

In 2016, the university of South Dakota led by Dongming Mei and Jing Liu, a former student of Iris, started a Program for International Research and Education (PIRE) funded by the US National Science Foundation in 2016.  They formed an international collaboration including some universities in the US, China and elsewhere, and Iris' group at the MPI [https://pire.gemadarc.org].
Tsinghua university was part of this collaboration and in 2018, they organized the first collaboration meeting at Xichang with a visit to CJPL, followed by a school organized in Chengdu by Sichuan university. 
In 2019 such a meeting was organized by Iris and her excellent secretary, Syblle Rodriguez, at the MPI. Iris found the teaching at such schools inspiring. 

The contact with Tsinghua and Sichuan university also persisted 
connected with the Chinese JinPing underground Laboratory (CJPL). Iris stayed involved in the development 
of the second phase of CJPL.  She has learned some Chinese and
enjoyed a lot of travel in South West China, often together with
her old friend Xiang and students. The visits to CJPL were of course central to her travels. Figures\,\ref{fig:CJPL1} and\,\ref{fig:CJPL2} give some impressions from 2018 and 2025, respectively. In 2018, the late halls had been completed. In 2025, the interior was almost finished and a lot of equipment had moved in. However, there is still space for many more experiments.

\begin{figure}[!tb]
    \centering
    \includegraphics[width=0.45\textwidth]{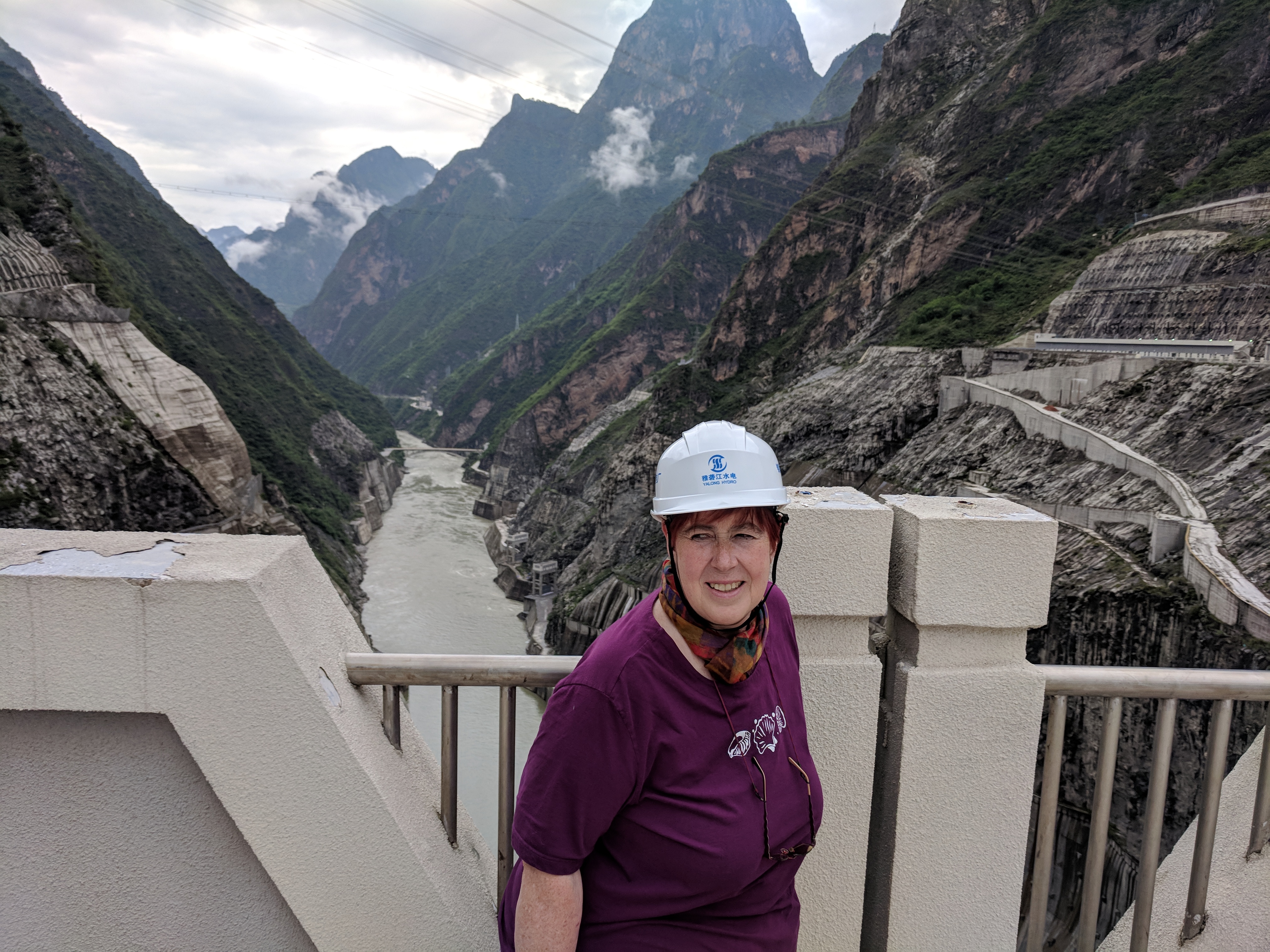} 
    \includegraphics[width=0.45\linewidth]{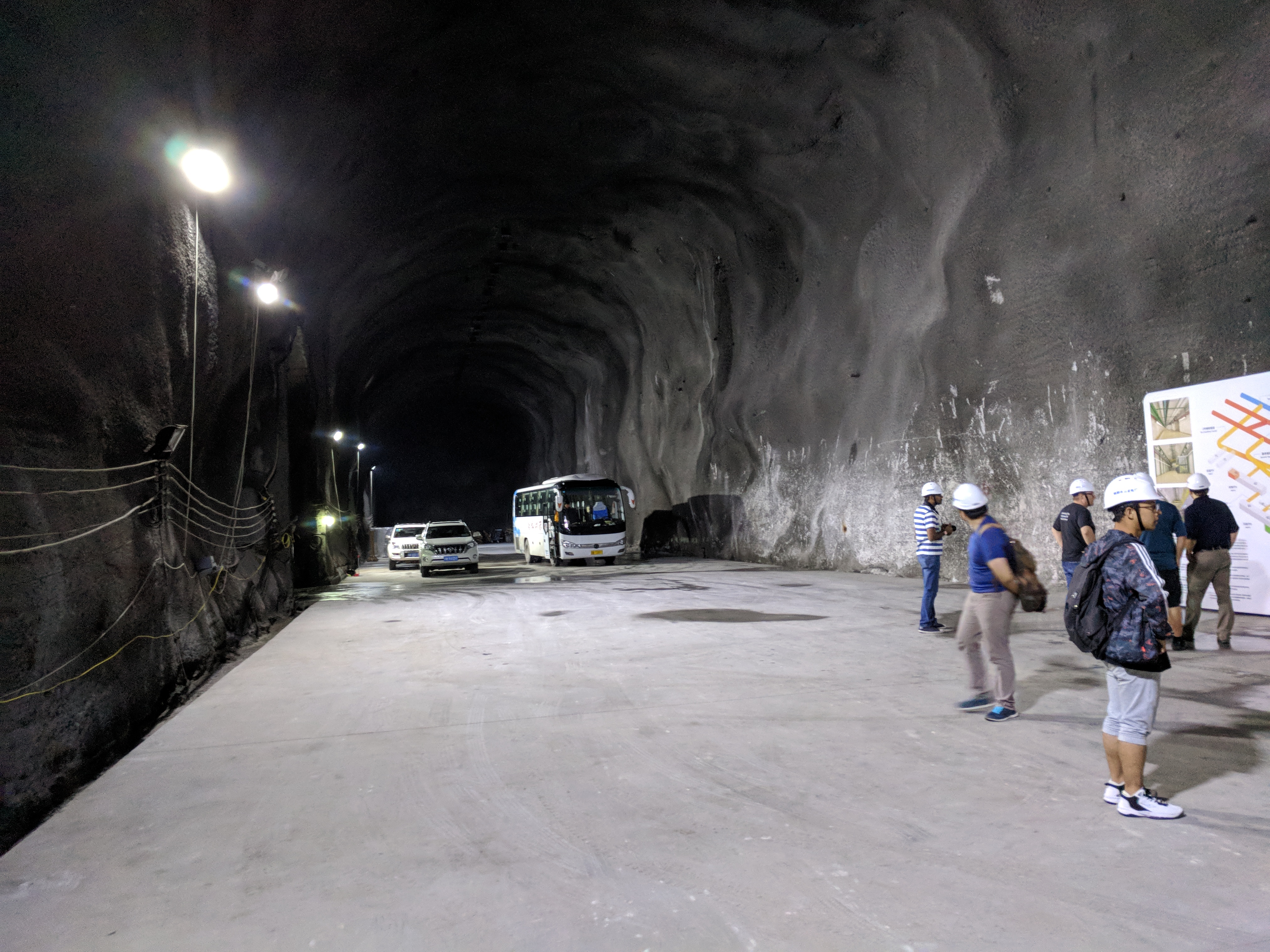}
    \includegraphics[width=0.35\textwidth]{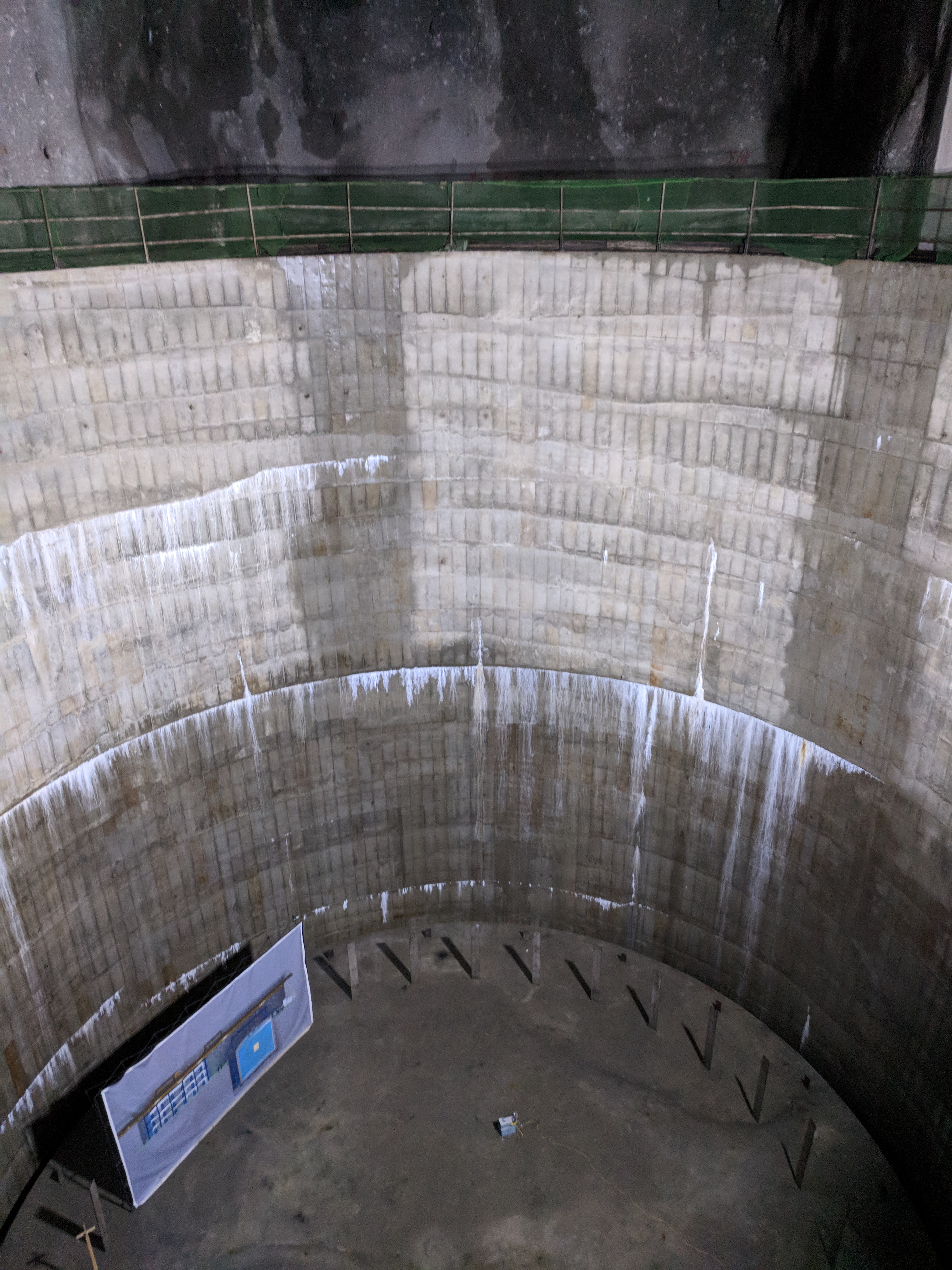} 
    \includegraphics[width=0.35\linewidth]{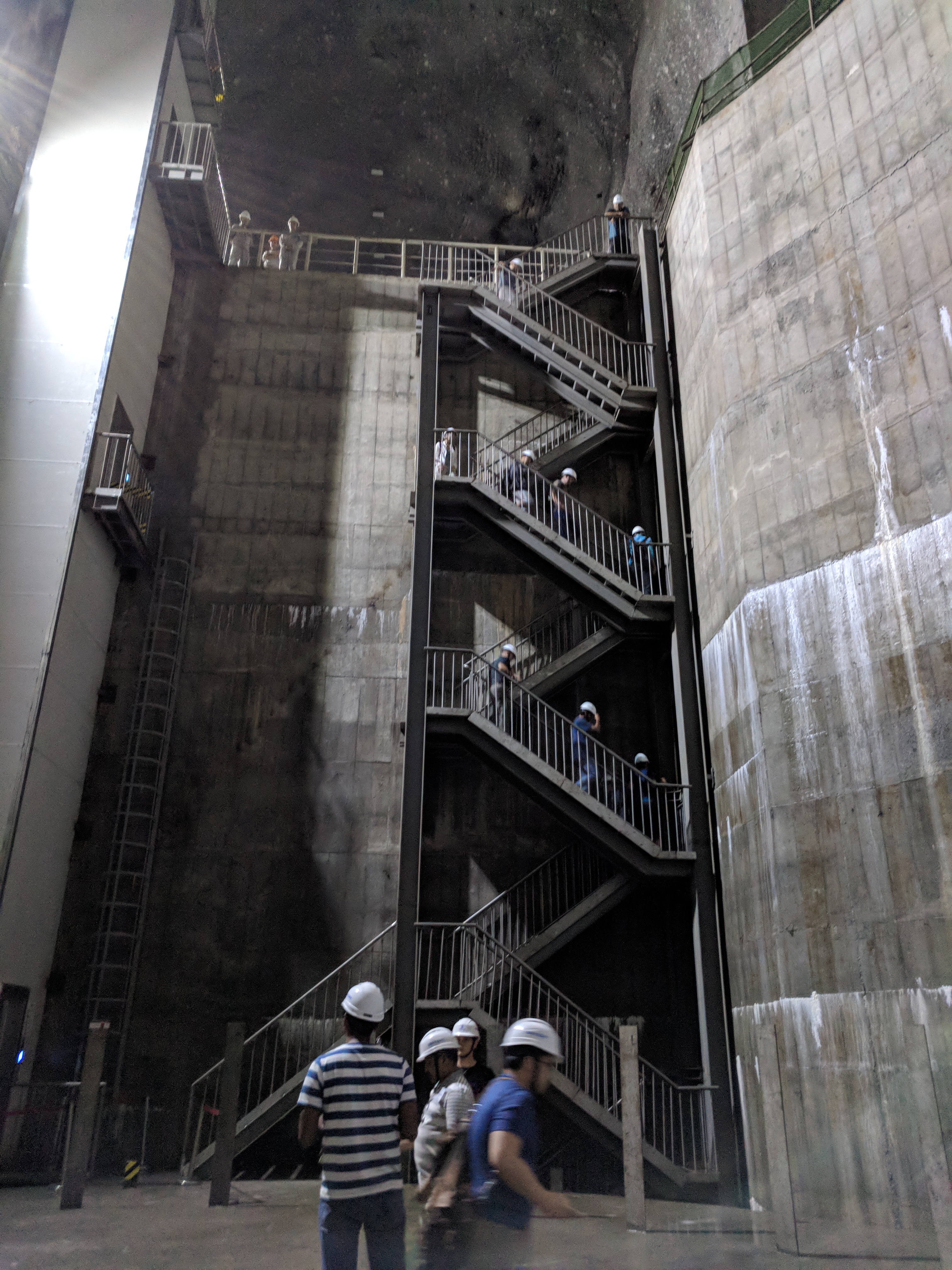}   
    \caption{CJPL is built into the support tunnels of the Jinping II hydropower plant [https://infra.global/projects/jinping-ii-hydropower-station/]. Iris is standing on the 305\,m high dam.
    The CJPL II laboratory has four 124\,m long halls. One of the halls has an extra pit fo 18\,m depth for the China Dark matter experiment. [Credit: Iris Abt, 2018] }
    \label{fig:CJPL1}
\end{figure}

\begin{figure}[!tb]
    \centering
    
    \includegraphics[width=0.45\linewidth]{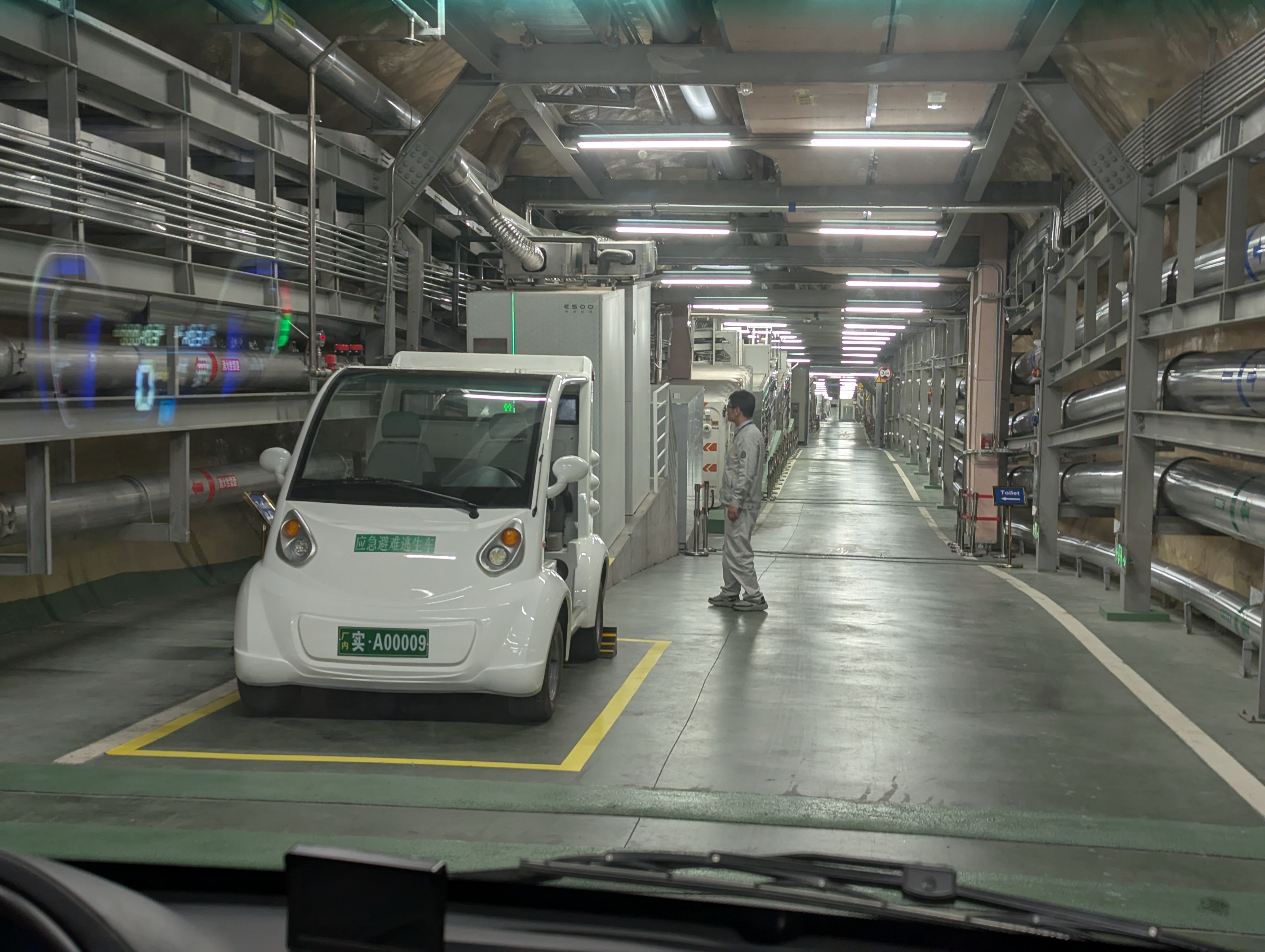}
    \includegraphics[width=0.45\textwidth]{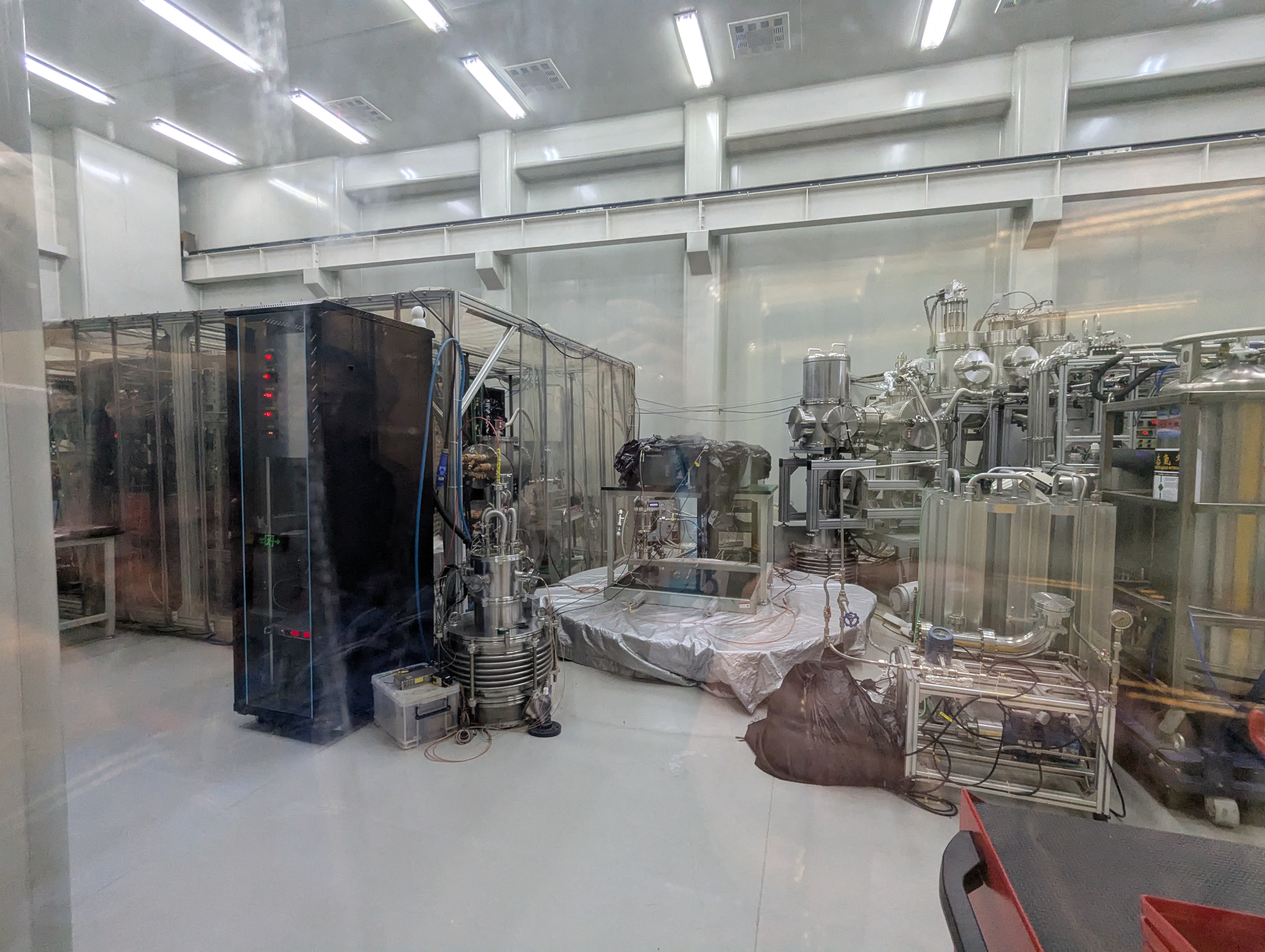} 
    \includegraphics[width=0.35\textwidth]{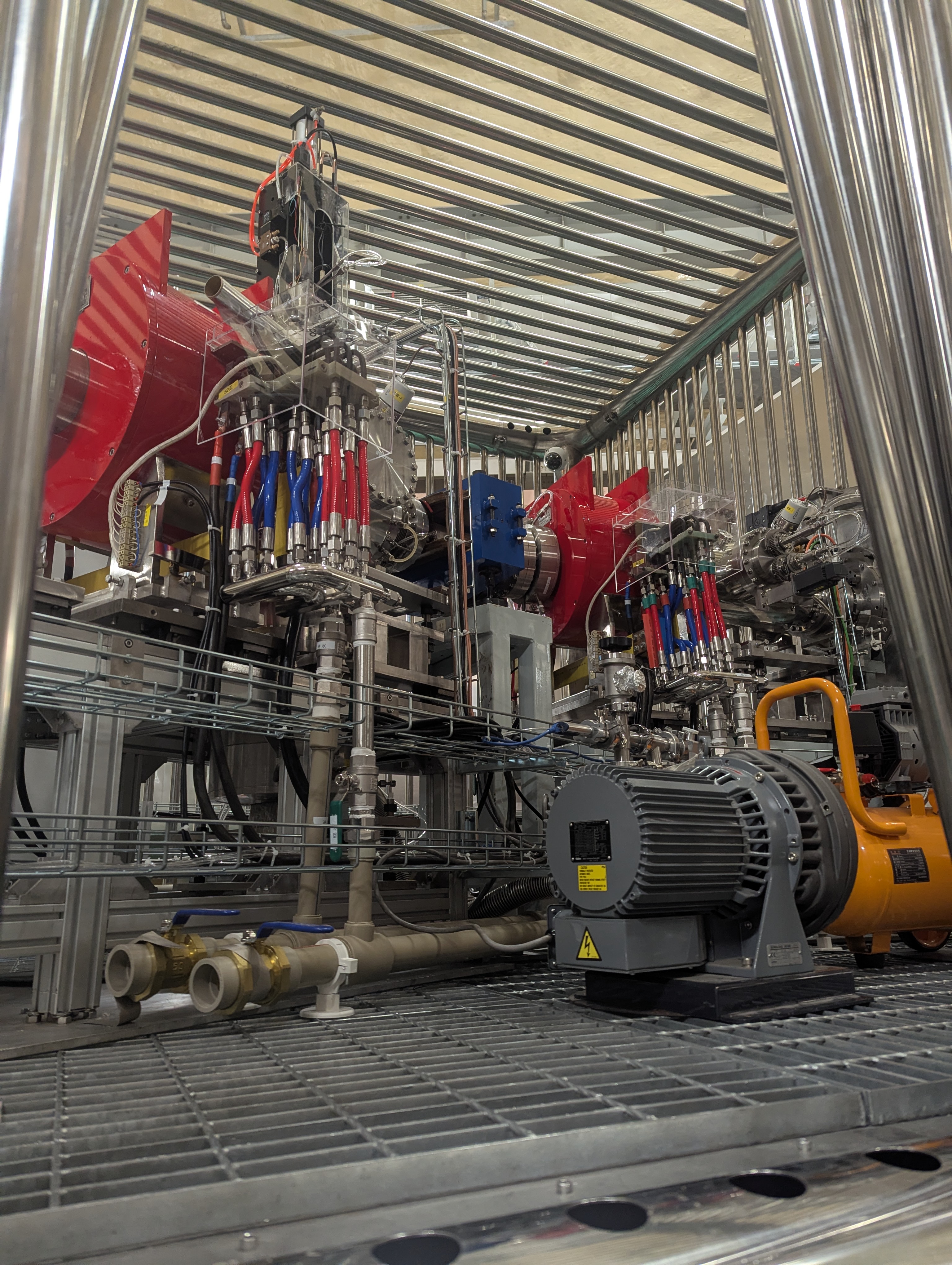} 
    \includegraphics[width=0.35\linewidth]{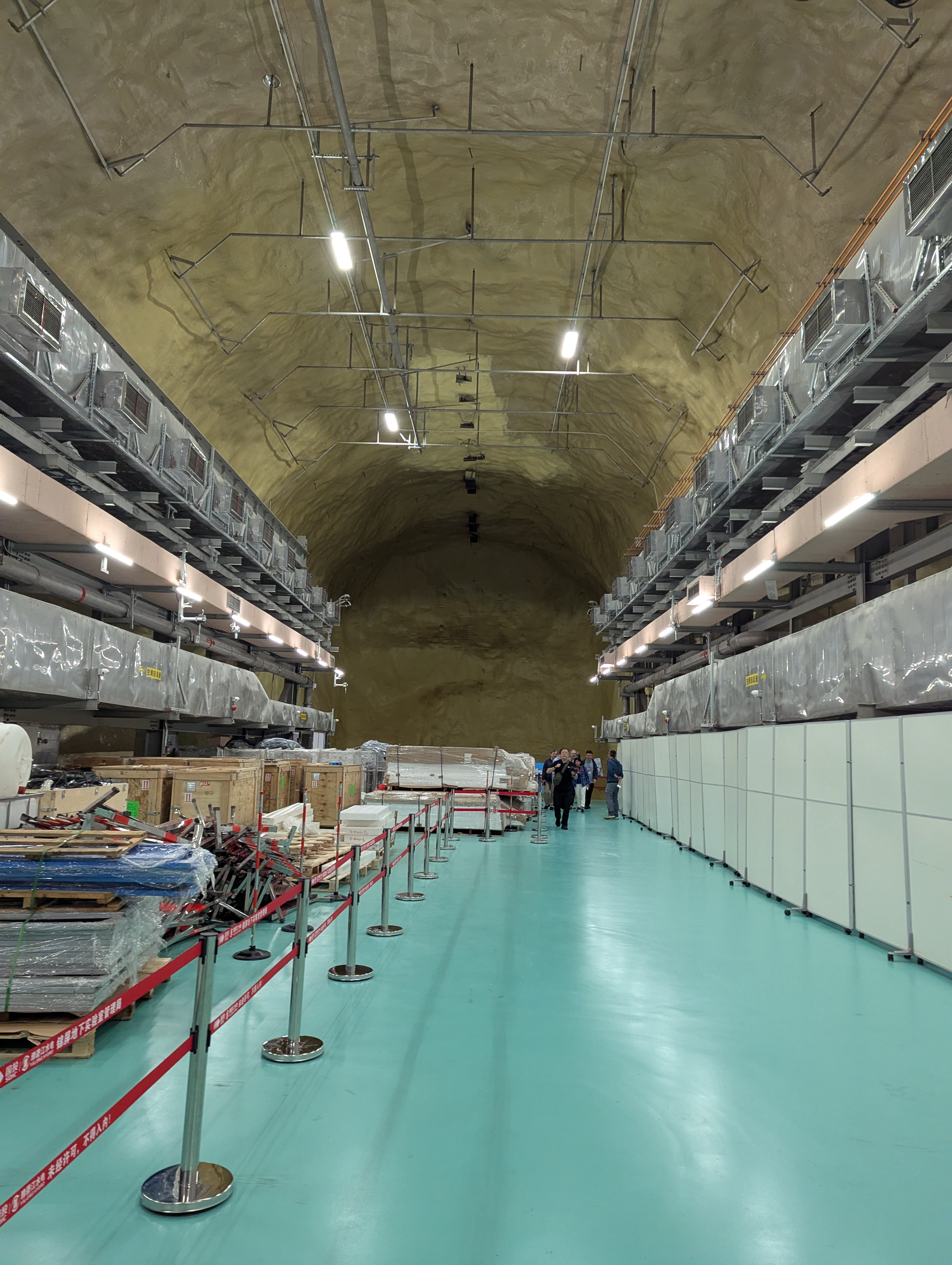}   
    \caption{The corridors of CJPL seem endless. A lot of equipment has moved into the halls but there is plenty of space left.
    [Credit: Iris Abt, 2025]}
    \label{fig:CJPL2}
\end{figure}

As a retiree Iris is presently still a member of the CJPL International
Advisory panel and serves as an external expert to the National Chinese Science Foundation concerning the selection of new projects
for CJPL.

\section{Final words}

My answer to Iris after the interview to {\emph{I was always in the wrong place and at the wrong moment}} would be that we all did our own small part in contributing to experimental particle physics. Indeed after these pleasant three hours of interview,  she sent me a long and impressive list of accomplishments in her forty years in science. 

Her suggestion to the early career researchers? 
It is a demanding career path and one needs the compassion for research to overcome all the obstacles. And compassion is often not enough. You need to make an impression and you need luck and probably elbows. 
Especially to women: while career prospects and conditions are slowly changing for the better, there are still many challenges related to family, the need to move often to different countries or laboratories and the difficulty to find permanent positions in academia. But women should remember that it is also difficult for men.  
Iris Abt herself was equal opportunity officer at the MPI for physics for twenty years and could witness many of these difficulties for women in particular.  
{\emph{The good thing in basic research is that nobody expects you to be in office before 10 and you most of the time can manage your time pretty freely. And it provides the opportunity to meet people in many places,}} says Iris. 

Her final words are for her students and postdocs: She is proud of (and thankful to) all of them. They all made very good careers in academia or industry.

\section{Acknowledgments}

This work is partly funded by the Deutsche Forschungsgemeinschaft (DFG, German Research Foundation) under Germany’s Excellence Strategy – EXC 2121 „Quantum Universe“ – 390833306.

\printbibliography

\end{document}